\documentclass[%
 reprint,
 amsmath,amssymb,
 aps,
]{revtex4-2}

\usepackage{graphicx}
\usepackage{dcolumn}
\usepackage{bm}
\usepackage{soul}
\usepackage{comment}
\usepackage{xcolor}
\usepackage{subcaption} 
\usepackage[percent]{overpic}
\usepackage{caption}
\usepackage{MnSymbol}
\usepackage{cancel}
\usepackage{bbm}
\usepackage[normalem]{ulem}

\begin{document}

\preprint{APS/123-QED}

\title{Collective modes in non-Hermitian fermionic superfluids}


\author{Gabriel Angelo V. Vila}
 \altaffiliation[Also at ]{National Institute of Physics, University of the Philippines Diliman, Philippines.}
\author{Kristian Hauser Villegas}%
 \email{kvillegas@nip.upd.edu.ph}
\affiliation{%
 National Institute of Physics, University of the Philippines Diliman, Philippines.\\
}%

\date{\today}

\begin{abstract}The Higgs and Nambu–Goldstone modes are paradigmatic collective excitations in superconductors and superfluids. These modes are commonly analyzed within the pseudospin formulation of BCS theory, where the dynamics are obtained from the Heisenberg equations of motion for pseudospins. However, this construction becomes inconsistent when directly extended to non-Hermitian systems. In this work, we develop a consistent pseudospin framework for non-Hermitian fermionic superfluids based on the metricized formulation of non-Hermitian quantum mechanics. We apply this formalism to a driven non-Hermitian BCS-type Hamiltonian with complex pairing interaction and analyze its collective excitation spectrum. We find that, in addition to the conventional Higgs (amplitude) mode, the system hosts a novel phase mode that has no counterpart in Hermitian superfluids. Remarkably, this mode is gapped even in the absence of the Anderson–Higgs mechanism, as is the case in neutral superfluids. Furthermore, the resonance spectrum depends explicitly on the initial nongauge phase of the complex order parameter and the dynamical response remains finite at resonance, in contrast to the divergence characteristic of Hermitian systems. These resonances disappear upon the emergence of exceptional points.

\end{abstract}

\maketitle


\section{Introduction}
\label{sec:intro}
Non-Hermitian Hamiltonians provide a natural framework for describing open quantum systems, where coupling to an environment leads to gain and loss. In such effective descriptions, complex energy eigenvalues encode both coherent dynamics and non-unitary processes arising from environmental interactions. As a result, non-Hermitian quantum mechanics extends the conventional Hermitian paradigm and captures environment effects that are ubiquitous in realistic quantum systems.

Beyond modeling openness, non-Hermitian systems exhibit qualitatively new phenomena without Hermitian analogues, including the appearance of exceptional points \cite{HeissSteeb1991, Dembowski2001, Heiss2012} and the breakdown of conventional bulk–boundary correspondence due to the non-Hermitian skin effect \cite{Dembowski2001, Gong2018, Lin2020}. Experimentally, non-Hermitian physics has been realized across diverse platforms, including ultracold atoms \cite{Takasu2020, Chen2022}, electrical and superconducting circuits \cite{Schindler2011, Helbig2020, Naghiloo2018, Ashida2020}, exciton-polariton condensates \cite{Deng2010, Gao2015}, and other systems. At the same time, non-Hermitian quantum mechanics has motivated foundational advances, such as biorthogonal \cite{Brody2014} and metricized formulations of quantum theory \cite{Mostafazadeh2004, Mostafazadeh2018, Ju2019}.

Recently, non-Hermitian quantum mechanics—particularly the biorthogonal formalism—has been applied to the study of non-Hermitian fermionic superfluids and superconductors \cite{Ghatak2018, Yamamoto2019}. This line of research has uncovered a range of novel and counterintuitive phenomena without Hermitian counterparts. Representative examples include reentrant superfluidity driven by increasing dissipation, dissipation-enhanced superfluid gaps \cite{Yamamoto2019}, and a paramagnetic Meissner response \cite{Tamura2025}. Moreover, the interplay between the non-Hermitian skin effect and the emergence of localized Majorana zero modes has been investigated in non-Hermitian second-order topological superconductors \cite{Ghosh2022, Ji2025}.

A key characteristic of the fermionic superfluid phase is the presence of the Higgs amplitude and Nambu-Goldstone collective modes as a result of spontaneous symmetry breaking. Experimentally, these collective modes have been realized and observed using light spectroscopy systems \cite{Measson2014, Matsunaga2013}, Bragg spectra \cite{Hoinka2017Goldstone}, and point contacts \cite{Uchino2020NGPointContact}. They are also realized using ultracold atom systems, which has the additional advantage of being tunable \cite{Vidanovic2014, Barontini2013, Tomita2017}. One such setup has allowed for the observation of the Higgs mode in fermionic superfluids \cite{Behrle2018}. However, these systems involve dissipative inelastic scattering. As such, realizations of superfluidity using ultracold atoms have been suggested to be naturally modeled by a non-Hermitian BCS Hamiltonian with a complex-valued interaction and using the biorthogonal framework \cite{Yamamoto2019}.

Two immediate challenges arise in the study of collective modes in non-Hermitian superconductors and superfluids. First, in Hermitian systems, the Anderson pseudospin formalism provides a simple, powerful, and widely used framework for describing collective excitations. This formalism, however, relies on pseudospin equations of motion derived from the Heisenberg equations for the corresponding pseudospin operators. In non-Hermitian systems, the conventional Heisenberg picture is not generally consistent with the corresponding Schr\"{o}dinger picture and, moreover, operator commutation and anticommutation relations are not preserved under time evolution \cite{Ju2025}. These issues will be discussed further in Sec.~\ref{sec:pseudospin}. Second, there is evidence using numerical simulations that the standard biorthogonal formalism does not provide an adequate description of quantum dynamics for time-dependent non-Hermitian Hamiltonians \cite{Sim2025}, which are essential for describing driven collective modes.

In this work, we formulate the pseudospin formalism within the Hilbert-space metricized framework \cite{Ju2025}, which generalizes the conventional biorthogonal approach. This formalism provides a consistent extension of the Heisenberg picture to non-Hermitian systems while maintaining full compatibility with the Schrödinger picture \cite{Ju2025}. Using this metricized pseudospin framework, we then investigate the collective modes of a non-Hermitian fermionic superfluid. 

The Nambu-Goldstone mode in dissipative superfluids has previously been studied within the framework of the Lindblad master equation \cite{Yamamoto2021}. Here, by contrast, we demonstrate the emergence of a novel phase mode that is intrinsic to non-Hermitian pairing dynamics and fundamentally distinct from the conventional Nambu-Goldstone mode. Moreover, unlike the gapless Nambu-Goldstone excitation, this mode remains gapped even in neutral fermionic superfluids.

\section{Non-Hermitian BCS Theory in pseudospin formalism}
\label{sec:pseudospin}
Similar to the Hermitian case, our starting point is the reduced Bardeen–Cooper–Schrieffer (BCS) Hamiltonian
\begin{align}
\label{eq:hamiltonian1}
    H=\sum_{\mathbf{k},\sigma}\varepsilon_{\mathbf{k}-\mathbf{A}(t)}c^\dagger_{\mathbf{k}\sigma}c_{\mathbf{k}\sigma}-U\sum_{\mathbf{k},\mathbf{p}}c^\dagger_{\mathbf{k}\uparrow} c^\dagger_{-\mathbf{k}\downarrow}c_{-\mathbf{p}\downarrow}c_{\mathbf{p}\uparrow},
\end{align}
where $c_{\mathbf{k}\sigma}$ and $c^\dagger_{\mathbf{k}\sigma}$ are the fermionic annihilation and creation operators, $\mathbf{A}(t)$ is an effective vector potential from external driving, and $U=U_0+i\gamma$ is the complex-valued interaction strength, so that this Hamiltonian is non-Hermitian. The coupling to a time-dependent vector potential $\mathbf{A}(t)=\mathbf{A}_0 \sin (\Omega t)$ was introduced via the Peierls substitution $\xi_\mathbf{k} \rightarrow \xi_{\mathbf{k} - \mathbf{A}(t)}$ \cite{Tsuji2015}. This model is the effective Hamiltonian in the no-jump limit of the Lindblad master equation for fermionic superfluids with two-body losses, with dissipation strength $2 \gamma$ \cite{Yamamoto2019}.

Since we are dealing with a non-Hermitian Hamiltonian, we need to redefine how the expectation values are to be computed. The conventional way is to use biorthogonal quantum mechanics, as previously done in non-Hermitian BCS superfluids \cite{Yamamoto2019}. However, due to external driving $\mathbf{A}(t)$, a more appropriate framework is the metricized formalism, which can be viewed as a natural generalization of the biorthogonal formalism when time-dependence is involved \cite{Sim2025}.

\subsection{Review of metricized non-Hermitian quantum mechanics.}
In the Heisenberg picture of Hermitian quantum mechanics, the evolution of an operator $\mathcal{O}$ is given by
In the Heisenberg picture of Hermitian quantum mechanics, the evolution of an operator $\mathcal{O}$ is given by
In the Heisenberg picture of Hermitian quantum mechanics, the evolution of an operator $\mathcal{O}$ is given by
In the Heisenberg picture of Hermitian quantum mechanics, the evolution of an operator $\mathcal{O}$ is given by
\begin{align}
\label{operatorevolve}
    \mathcal{O}_H(t)=U^\dagger(t,t_0)\mathcal{O}_SU(t,t_0),
\end{align}
where $U(t,t_0)$ is the evolution operator.

In non-Hermitian quantum mechanics, this evolution is no longer a similarity transformation since
\begin{align}
    U^\dagger(t,t_0)\neq U^{-1}(t,t_0).
\end{align}

Consequently, the equivalence between the Heisenberg and Schr\"{o}dinger pictures breaks down. More seriously, the canonical commutation relations satisfied by operators are generally not preserved under time evolution \cite{Ju2025}.

These problems can be resolved by the metricized formalism where a Hilbert space metric $G(t)$ is introduced \cite{Ju2025}. From the usual ket $|\psi\rangle$ and bra $\langle\psi |$, one defines a new ket and its dual using this metric
\begin{align}
\label{eq:metricket}
|\psi(t)\rrangle &\equiv | \psi(t)\rangle\\
\label{eq:metricdual}
\llangle \psi(t)| &\equiv \langle \psi(t)| G(t),
\end{align}
in analogy with the covariant and contravariant vectors in differential geometry.

The inner product is then written as
\begin{align}
    \llangle\psi_1(t) | \psi_2(t) \rrangle = \langle \psi_1(t) | G(t) | \psi_2(t)\rangle. 
\end{align}

Requiring that the metricized inner product $\llangle\psi(t)|\psi(t)\rrangle$ be time independent gives us the equation that must be obeyed by the metric \cite{Ju2019}
\begin{align}
\label{eq:metricequation}
     \partial_tG(t)=i[G(t)H_S(t)-H_S^\dagger(t)G(t)],
\end{align}
where $H_S(t)$ is the Hamiltonian in the Schr\"{o}dinger picture—the $t$ in its argument means explicit time dependence.


By taking the time derivative of Eqs. \eqref{eq:metricket} and \eqref{eq:metricdual}, then using Eq. \eqref{eq:metricequation} and the fact that the original ket obeys the conventional Schr\"{o}dinger equation
\begin{align}
i\partial_t|\psi\rangle=H_S(t)|\psi\rangle,
\end{align}
it can be shown that the time evolution of these metricized ket and its dual are given by
\begin{align}
   \partial_t |\psi(t)\rrangle &= -iH_S(t) |\psi(t)\rrangle \\
    \partial_t \llangle\psi(t)| &= i \llangle \psi(t)| H_S(t).
\end{align}

The formal solutions can be written in terms of time evolution operator
\begin{align}
    |\psi(t)\rrangle = U(t,t_0) | \psi(t_0)\rrangle \\
    \llangle \psi(t) | = \llangle \psi(t_0) |U^{-1}(t, t_0).
\end{align}

Note that the metric dual is evolved via the inverse $U^{-1}(t, t_0)$ instead of the usual Hermitian conjugate $U^\dagger(t, t_0)$.

Consequently, a Heisenberg-like representation can be constructed, with the operators obeying the equation of motion
\begin{align}
\label{eq:heisenberg_metric}
    \frac{d\mathcal{O}_H(t)}{dt}=i[H_H(t), \mathcal{O}_H(t)] + [\partial_t \mathcal{O}_S(t)]_H,
\end{align}
where the subscript $H$ indicates that $O_H(t)$ is the Heisenberg representation of operator $\mathcal{O}$, and is related to its Schr\"odinger representation via the evolution
\begin{align}
\label{eq:operatorevolution}
    \mathcal{O}_H(t)=U^{-1}(t,t_0)\mathcal{O}_S(t)U(t,t_0).
\end{align}

Note that the right-hand side involves the operator $U^{-1}(t,t_0)$ rather than $U^\dagger(t,t_0)$. Consequently, the time evolution of an operator is given by a similarity transformation. This guarantees that the Heisenberg formalism defined here is consistent with the Schr\"{o}dinger picture and, in addition, the commutation or anti-commutation relations satisfied by operators are preserved under time evolution \cite{Ju2025}.

The expectation value of an observable in the Heisenberg picture $\mathcal{O}_H(t)$ with respect to the right state $|\psi\rangle$ is computed as
\begin{align}
\label{eq:observableexpectation}
\langle \mathcal{O}\rangle=\llangle\psi|\mathcal{O}_H(t)|\psi\rrangle=\langle\psi|G(t)\mathcal{O}_H(t)|\psi\rangle.
\end{align}

We next employ this metricized Hilbert-space formulation to derive the appropriate pseudospin equations of motion together with the corresponding self-consistency equations.

\subsection{Mean field Hamiltonian.}
The mean-field BCS theory requires the introduction of an order parameter, which is the expectation value of the pairing operator. Due to non-Hermiticity of the Hamiltonian Eq. \eqref{eq:hamiltonian1}, performing a mean-field approximation requires careful definition of the order parameter that reflects the correct prescription of expectation values, which is given by Eq. \eqref{eq:observableexpectation} in the case of Hilbert-space metricized formulation. As we will now show, there are two order parameters $\Delta(t)$ and $\bar{\Delta}(t)$ due to non-Hermiticity \cite{Yamamoto2019}, which are time-dependent due to the external driving. They are defined using the metricized inner product as
\begin{align}
    \label{eq:scemetricized1a}
    \Delta(t)\equiv&-U\sum_\mathbf{k}\llangle \text{BCS}(t)| c_{-\mathbf{k}\downarrow}c_{\mathbf{k}\uparrow}|\text{BCS}(t)\rrangle\\
    \label{eq:scemetricized1b}
    \bar{\Delta}(t)\equiv&-U\sum_\mathbf{k}\llangle \text{BCS}(t)|  c_{\mathbf{k}\uparrow}^\dagger c_{-\mathbf{k}\downarrow}^\dagger|\text{BCS}(t)\rrangle ,
\end{align}
where $|\text{BCS}(t)\rrangle \equiv |\text{BCS}(t) \rangle$ and $\llangle \text{BCS}(t) | \equiv \langle \text{BCS}(t) |G(t)$ are the metricized ket and dual states for the time-dependent non-Hermitian BCS ground state. Note that $\langle \text{BCS}(t) |$ is the dual bra of the \textit{right} ground state $|\text{BCS}(t) \rangle$, as prescribed by Eq. \eqref{eq:metricdual}. We further emphasize that, in contrast to the Hermitian BCS theory, $\Delta(t)$ and $\bar{\Delta}(t)$ are not Hermitian conjugates due to the complex-valued interaction $U$ and the structure of the inner product. As we will see, this will lead to distinct characteristics of non-Hermitian fermionic superfluids compared to their Hermitian counterparts.

Using the mean-field approximation to the Hamiltonian Eq. \eqref{eq:hamiltonian1} leads to the Bogoliubov-de Gennes form
\begin{align}
    \label{eq:BdG_NH_driving}
    H_S (t)=&\sum_\mathbf{k}(c^\dagger_{\mathbf{k}\uparrow} \; c_{-\mathbf{k}\downarrow})
    \begin{pmatrix}
    \xi_{\mathbf{k}-\mathbf{A}(t)} & \Delta(t)\\
    \bar{\Delta}(t) & -\xi_{\mathbf{k}+\mathbf{A}(t)}
    \end{pmatrix}
    \begin{pmatrix}
        c_{\mathbf{k}\uparrow}\\
        c^\dagger_{-\mathbf{k}\downarrow}
    \end{pmatrix}\\
    \label{bdghamiltonian1}
    =&\sum_\mathbf{k}\psi^\dagger_\mathbf{k}H_{BdG}\psi_\mathbf{k} ,
\end{align}
where we introduced the Nambu spinor
\begin{align}
    \psi_\mathbf{k}=(c_{\mathbf{k}\uparrow},c^\dagger_{-\mathbf{k}\downarrow})^T
\end{align}
and the Bogoliubov-de Gennes Hamiltonian
\begin{align}
\label{eq:BdGhamiltonian}
    H_{BdG}=
    \begin{pmatrix}
        \xi_{\mathbf{k}-\mathbf{A}(t)} & \Delta(t) \\
        \bar{\Delta}(t) & -\xi_{\mathbf{k}+\mathbf{A}(t)}
    \end{pmatrix}.
\end{align}

We now define the pseudospin operator \cite{Tsuji2015}
\begin{align}
    \label{pseudospin}
    \hat{\boldsymbol{\sigma}}_{\mathbf{k}, S}\equiv\frac{1}{2}\psi^\dagger_{\mathbf{k}}\boldsymbol{\tau}\psi_{\mathbf{k}},
\end{align}
where $\{\tau^i|i=1,2,3\}$ are the Pauli matrices in particle-hole space. The BdG Hamiltonian, Eq. \eqref{eq:BdG_NH_driving}, written in terms of pseudospin operators becomes
\begin{align}
    \label{eq:NH_bcs_pseudospin}
    H_S (t)=\sum_\mathbf{k}2\mathbf{B}_\mathbf{k}(t) \cdot \hat{\boldsymbol{\sigma}}_{\mathbf{k}, S},
\end{align}
where the components of the pseudomagnetic field are calculated as
\begin{align}
    B^x_\mathbf{k}(t)=&\frac{1}{2}\text{Tr}(H_{BdG}\tau^x)=\frac{1}{2}\big(\Delta(t)+\bar{\Delta}(t)\big) \label{eq:Bx}\\
    B^y_\mathbf{k}(t)=&\frac{1}{2}\text{Tr}(H_{BdG}\tau^y)=\frac{i}{2}\big(\Delta(t)-\bar{\Delta}(t)\big) \label{eq:By}\\
    B^z_\mathbf{k}(t)=&\frac{1}{2}\text{Tr}(H_{BdG}\tau^z)=\frac{1}{2}(\xi_{\mathbf{k}-\mathbf{A}(t)}+\xi_{\mathbf{k}+\mathbf{A}(t)}) \label{eq:Bz} .
\end{align}
Due to the complex interaction, the pseudospin formalism for non-Hermitian superfluids involves complex pseudomagnetic fields, in contrast to real pseudomagnetic fields for Hermitian superfluids.

The order parameters $\Delta(t)$ and $\bar{\Delta}(t)$ are not independent due to the global U(1) symmetry of the non-Hermitian Hamiltonian, Eq. \eqref{eq:hamiltonian1}. Specifically, if the non-Hermitian mean-field Hamiltonian satisfies the symmetry $H^\dagger = H^*$, the zeroth-order order parameters can be expressed as $\Delta^{(0)} = \Delta_{0} e^{i\theta^{(0)}}$ and $\bar{\Delta}^{(0)} = \Delta_{0} e^{-i\theta^{(0)}}$, where $\Delta_0 \in \mathbb{C}$ \cite{Yamamoto2019}. We emphasize that $\Delta_0$ is complex in contrasts to the Hermitian case where it is real. To zeroth order, the order parameter can therefore be written as
\begin{align}
    \Delta^{(0)}=& |\Delta^{(0)}| e^{i(\phi^{(0)}+\theta^{(0)})}\\
     \bar{\Delta}^{(0)}=& |\Delta^{(0)}| e^{i(\phi^{(0)}-\theta^{(0)})},
\end{align}
where $|\Delta^{(0)}|$ is the amplitude, $\theta^{(0)}$ is the usual phase of the order parameter, and $\phi^{(0)}$ is a new phase that comes from the non-Hermiticity of the mean-field Hamiltonian \cite{Yamamoto2019}.

We can choose a gauge so that $\theta^{(0)}=0$ and we have
\begin{equation}
    \label{eq:delta_zero}
    \Delta^{(0)}=\bar{\Delta}^{(0)} = |\Delta^{(0)}| e^{i\phi^{(0)}}.
\end{equation}

In contrast to $\theta^{(0)}$, the non-Hermitian phase $\phi^{(0)}$ cannot be eliminated by a gauge transformation. It thus constitutes an independent parameter characterizing the non-Hermitian BCS ground state.

Fluctuations of the phase $\theta$ correspond to the Nambu-Goldstone mode, which is gapless in neutral superfluids and becomes gapped in superconductors through the Anderson-Higgs mechanism. As we will show below, fluctuations of the non-Hermitian phase $\phi$ give rise to a distinct collective mode that is gapped even in neutral superfluids.

\subsection{Exceptional points}
\label{subsec:exceptionalpoints}
One of the hallmarks of non-Hermitian Hamiltonians is the possible existence of the exceptional points where the eigenvectors coalesce in addition to the degeneracy in the eigenvalues \cite{Heiss2012}. Let us therefore discuss the apperance and properties of the exceptional points that appear in the Bogoliubov-de Gennes Hamiltonian Eq. \eqref{eq:BdGhamiltonian}. The excitation energies, including the hole-like energies, are
\begin{align}
\label{eq:excitation energies}
E_\mathbf{k}=\pm \sqrt{\xi^2_\mathbf{k}+\bar{\Delta}\Delta}=\pm\sqrt{\xi^2_\mathbf{k}+|\Delta_0|^2e^{2i\phi}}.
\end{align}

The exceptional points occur when 
\begin{align}
\label{eq:epenergy}
    \xi^{(EP)}_\mathbf{k}=\pm i|\Delta_0|e^{i\phi^{(0)}}.
\end{align}

Recall that the non-Hermiticity of our Hamiltonian originates from two-body losses, which give rise to an effective complex interaction $U$. By contrast, the single-particle bare dispersion $\xi_{\mathbf{k}}$ remains real, implying that the exceptional points occur in Eq. \eqref{eq:epenergy} occur when $\phi^{(0)}=\pm \pi/2$, where it gives $\xi_{\mathbf{k}}^{(\mathrm{EP})}=\pm |\Delta_0|$. In momentum space, these exceptional points form two surfaces located at $\pm |\Delta_0|$ away from the Fermi surface.

The right and left eigenvectors corresponding to the two exceptional points above are
\begin{align}
\label{eq:rightepvectors}
|R^+\rangle=(+i,1)^T,\;\;
|R^-\rangle=(-i,1)^T
\end{align}
and 
\begin{align}
|L^+\rangle=(+i,1),\;\;
|L^-\rangle=(-i,1),
\end{align}
respectively.

Since the two components of the eigenvectors of the Bogoliubov-de Gennes Hamiltonian correspond to the particle and hole degrees of freedom, it is natural to ask whether the coalesced eigenvectors form Majorana states. To address this question, we consider the right eigenvectors in Eq.~\eqref{eq:rightepvectors}.
 Applying the charge conjugation operator gives
\begin{align}
C|R^{\pm}\rangle=\tau^xK|R^{\pm}\rangle=|R^{\mp}\rangle,
\end{align}
which shows that the charge conjugation operation interchanges the two exceptional points. The eigenvectors corresponding to the exceptional points therefore do not form Majorana states.

We will further show below that the resonance for the collective modes disappear with the apperance of the exceptional points when $\phi^{(0)}=\pm \pi/2$.

\subsection{Pseudospin equation of motion and self-consistent equations.}
Using Eq. \eqref{eq:heisenberg_metric}, the equation of motion for the Heisenberg-picture pseudospin operator is
\begin{align}
\frac{d}{dt}\hat{\boldsymbol{\sigma}}_\mathbf{k}(t)=i[H_H(t),\hat{\boldsymbol{\sigma}}_\mathbf{k}(t)].
\end{align}
Here, the pseudospin operator without the subscript \lq\lq H\rq\rq is understood to be written in the Heisenberg picture. We can write the operators in the commutator above in terms of the Schr\"{o}dinger picture using Eq. \eqref{eq:operatorevolution}, then simplify to get
\begin{align}
i[H_H(t),\hat{\boldsymbol{\sigma}}_\mathbf{k}(t)]=&\ i [U^{-1}H_SU,U^{-1}\hat{\boldsymbol{\sigma}}_{\mathbf{k},S}U]\\
=&\ U^{-1}i[H_S,\hat{\boldsymbol{\sigma}}_{\mathbf{k},S}]U\\
=&\ U^{-1} 2\mathbf{B}_\mathbf{k}(t)\times\hat{\boldsymbol{\sigma}}_{\mathbf{k},S}U\\
=&\ 2\mathbf{B}_\mathbf{k}(t)\times\hat{\boldsymbol{\sigma}}_\mathbf{k}(t).
\end{align}

Hence, the pseudospin equation of motion takes the same form as their Hermitian counterpart
\begin{align}
\label{eq:eom1}
\frac{d}{dt}\hat{\boldsymbol{\sigma}}_\mathbf{k}(t)=2 \mathbf{B}_\mathbf{k}(t)\times\hat{\boldsymbol{\sigma}}_\mathbf{k}(t).
\end{align}
While they have similar forms, we emphasize that the pseudospin formalisms in Hermitian and non-Hermitian BCS theories are different as the evolution of operators and the definition of expectation values are different.

The solution to Eq. \eqref{eq:eom1} is subject to the self-consistent equations, Eqs. \eqref{eq:scemetricized1a} and \eqref{eq:scemetricized1b}. Written in terms of pseudospin operators, they become
\begin{align}
    \label{eq:scemetricized2a}
    \Delta(t)=&-U\sum_\mathbf{k}\llangle \text{BCS}(0)|[\hat{\sigma}^x_\mathbf{k}(t)-i\hat{\sigma}^y_\mathbf{k}(t)]|\text{BCS}(0)\rrangle \\
    \label{scemetricized2b}
    \bar{\Delta}(t)=&-U\sum_\mathbf{k}\llangle\text{BCS}(0)|[\hat{\sigma}^x_\mathbf{k}(t)+i\hat{\sigma}^y_\mathbf{k}(t)]|\text{BCS}(0)\rrangle.
\end{align}

For a semi-classical treatment, the pseudospin operators in equations of motion, Eq. \eqref{eq:eom1}, are replaced with their metricized expectation values. We define the expectation value of pseudospin operator in Heisenberg picture as the Pauli sigma vector without hat
\begin{align}
\label{eq:operatorexpectation}
    \boldsymbol{\sigma}_\mathbf{k}(t)\equiv\llangle\text{BCS}(0)|\hat{\boldsymbol{\sigma}}_\mathbf{k}(t)|\text{BCS}(0)\rrangle.
\end{align}
Our full working equations are now
\begin{align}
\label{eq:eom2}
\frac{d}{dt}\boldsymbol{\sigma}_\mathbf{k}(t)=2 \mathbf{B}_\mathbf{k}(t)\times\boldsymbol{\sigma}_\mathbf{k}(t),
\end{align}
along with the self-consistent equations
\begin{align}
\label{eq:scemetricized2a}
\Delta(t)=&-U\sum_\mathbf{k}[\sigma^x_\mathbf{k}(t)-i\sigma^y_\mathbf{k}(t)]\\
\label{eq:scemetricized2b}
\bar{\Delta}(t)=&-U\sum_\mathbf{k}[\sigma^x_\mathbf{k}(t)+i\sigma^y_\mathbf{k}(t)].
\end{align}

One may ask how the metric $G(t)$ enters the calculation. Note that we are interested only in the expectation values of the pseudospin operators [Eq.~\eqref{eq:operatorexpectation}], and the solution of Eq.~\eqref{eq:eom2} already incorporates the effects of $G(t)$, such that its explicit form is not required. By contrast, if one wishes to calculate expectation values in the Schr\"{o}dinger picture by solving for $|\psi(t)\rangle$, then the explicit form of $G(t)$ must be determined and used in Eq.~\eqref{eq:observableexpectation}.

\section{Solution}
Having established that the pseudospin formalism can be extended to non-Hermitian fermionic superfluids, we now solve Eqs.~\eqref{eq:eom2}, \eqref{eq:scemetricized2a}, and \eqref{eq:scemetricized2b}. We consider a weak perturbation $\mathbf{A}(t)$ and expand the pseudospins perturbatively up to second order, which is the leading order in $\mathbf{A}(t)$ that gives rise to collective oscillations \cite{Tsuji2015}. Explicitly, the perturbative expansion of the pseudospin vector is
\begin{align}
    \label{eq:pseudospin_perturb}
    \boldsymbol{\sigma}_\mathbf{k}(t) &= \boldsymbol{\sigma}^{(0)}_\mathbf{k} + \boldsymbol{\sigma}^{(1)}_\mathbf{k}(t) +
    \boldsymbol{\sigma}^{(2)}_\mathbf{k}(t) .
\end{align}
Here, the superscripts in parentheses in $\boldsymbol{\sigma}^{(n)}_{\mathbf{k}}$ denote the $n$th-order contribution in the perturbative expansion, which scales as the $n$th power of the drving amplitude $\sim |\mathbf{A}_0|^n$.

Similalry, the order parameters $\Delta$ and $\bar{\Delta}$ are expanded as
\begin{align}
    \label{eq:delta_perturb}
    \Delta(t) =\Delta^{(0)} +\Delta^{(1)}(t)+\Delta^{(2)}(t) \\
    \label{eq:deltabar_perturb}
    \bar{\Delta}(t) = \bar{\Delta}^{(0)}+\bar{\Delta}^{(1)}(t)+\bar{\Delta}^{(2)}(t).
\end{align}

The expansion of the $x$ and $y$ pseudomagnetic field components, given by Eqs. (\ref{eq:Bx}) and (\ref{eq:By}), are
\begin{align}
    \label{eq:Bx_perturb}
    B_\mathbf{k}^x &= \Delta^{(0)} + \frac{1}{2} [\Delta^{(1)}(t)+\bar{\Delta}^{(1)}(t)] + \frac{1}{2} [\Delta^{(2)}(t)+\bar{\Delta}^{(2)}(t)] \\
    \label{eq:By_perturb}
    B_\mathbf{k}^y &= \frac{i}{2} [\Delta^{(1)}(t) - \bar{\Delta}^{(1)}(t)] + \frac{i}{2} [\Delta^{(2)}(t) - \bar{\Delta}^{(2)}(t)] .
\end{align}

For the $z$ component of the pseudomagnetic field, Eq. (\ref{eq:Bz}), we use
\begin{align}
\label{eq:Bz_perturb}
    \frac{1}{2}(\xi_{\textbf{k}-\textbf{A}(t)}+\xi_{\textbf{k}+\textbf{A}(t)})=\xi_\textbf{k} + \sum_{i,j=1}^3 A^i(t)A^j(t)\partial_i\partial_j \xi_\textbf{k} ,
\end{align}
where $A^i(t)$ are the vector potential components and $\partial_i\partial_j\equiv \partial_{k_i} \partial_{k_j}$. For systems with equivalent crystallographic directions, the second term can be written as \cite{Tsuji2015}
\begin{equation}
    \label{eq:driving_perturb}
     \sum_{i,j=1}^3A^i(t)A^j(t)\ \partial_i \partial_j \xi_\textbf{k} = \frac{1}{3} \nabla^2_\mathbf{k} \xi_\mathbf{k} |\mathbf{A}(t)|^2.
\end{equation}
Moreover, we consider a band dispersion $\xi_\mathbf{k}$ that satisfies
\begin{equation}
    \label{eq:dispersion_laplacian}
    \frac{1}{3} \nabla^2_\mathbf{k} \xi_\mathbf{k} = \alpha_1 \xi_\mathbf{k}
\end{equation}
for some constant $\alpha_1$, since this term gives the leading contribution to the Higgs oscillation \cite{Tsuji2015}. An example of such a band dispersion is the tight binding dispersion $\xi_\mathbf{k} = -2 t_0 \sum_{i=1}^3 \cos(a k_i) - \varepsilon_F$, for hopping amplitude $t_0$ and lattice spacing $a$. 

We are primarily interested in the resonance spectrum of the collective modes. For this purpose, it is sufficient to consider a sinusoidal external driving 
\begin{equation}
    \label{eq:driving}
    \mathbf{A}(t)=\mathbf{A}_0 \sin (\Omega t).
\end{equation}

With these considerations, the $z$ component of the pseudomagnetic field has the expansion
\begin{equation}
    \label{eq:Bz_perturb}
    B_\mathbf{k}^z = \xi_\textbf{k} +  \alpha_1 |\mathbf{A}_0|^2 \sin^2(\Omega t) \xi_\mathbf{k}.
\end{equation}

We then arrive at the perturbative pseudospin equations of motion
\begin{align}
    \label{eq:eom_zero}
    \frac{d\boldsymbol{\sigma}_\mathbf{k}^{(0)}}{dt} =& 2 \mathbf{B_k}^{(0)} \times \boldsymbol{\sigma}_\mathbf{k}^{(0)} \\
    \label{eq:eom_first}
    \frac{d\boldsymbol{\sigma}_\mathbf{k}^{(1)}}{dt} =& 2 \mathbf{B_k}^{(0)} \times \boldsymbol{\sigma}_\mathbf{k}^{(1)} + 2 \mathbf{B_k}^{(1)} \times \boldsymbol{\sigma}_\mathbf{k}^{(0)} \\
    \label{eq:eom_second}
    \frac{d\boldsymbol{\sigma}_\mathbf{k}^{(2)}}{dt} =& 2 \mathbf{B_k}^{(0)} \times \boldsymbol{\sigma}_\mathbf{k}^{(2)} + 2 \mathbf{B_k}^{(1)} \times \boldsymbol{\sigma}_\mathbf{k}^{(1)}(t)\nonumber\\
    &+ 2 \mathbf{B_k}^{(2)}(t) \times \boldsymbol{\sigma}_\mathbf{k}^{(0)} ,
\end{align}
where each of the \emph{n}th order pseudospin solutions $\boldsymbol{\sigma_k}^{(n)}$ is subject to the self-consistent equations
\begin{align}
    \label{eq:nscemetricized1a}
    \Delta^{(n)}(t)&=-U\sum_\mathbf{k}[\sigma^{(n)x}_\mathbf{k}(t)-i\sigma^{(n)y}_\mathbf{k}(t)]\\
    \label{eq:nscemetricized1b}
    \bar{\Delta}^{(n)}(t)&=-U\sum_\mathbf{k}[\sigma^{(n)x}_\mathbf{k}(t)+i\sigma^{(n)y}_\mathbf{k}(t)].
\end{align}

\begin{figure*}[!ht]
  \centering  
  {\makebox[0.325\textwidth]{\includegraphics[width=0.325\textwidth]{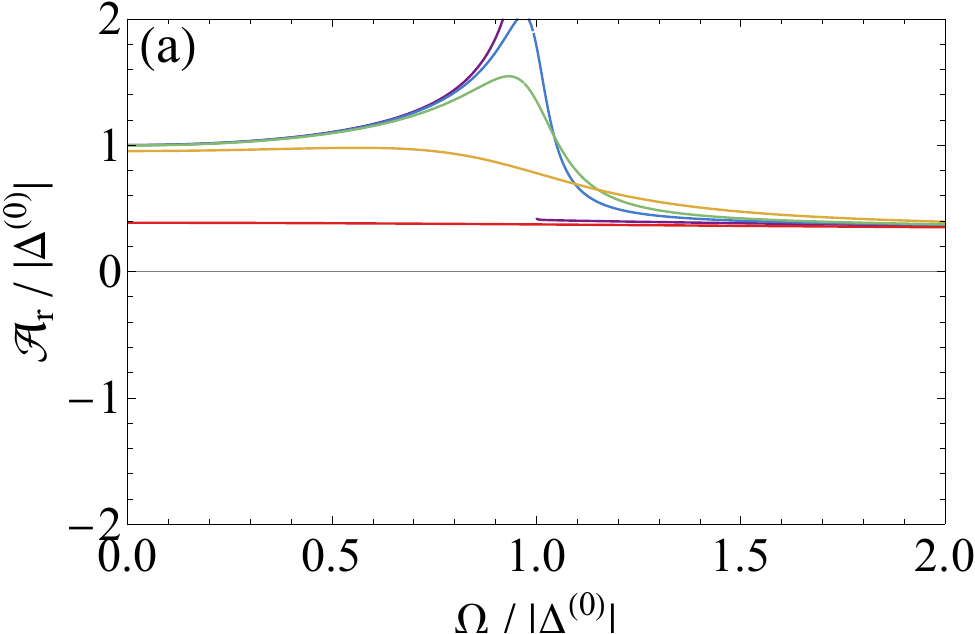}}\label{fig:RadPhiVaried}}
  \hfill
  {\makebox[0.325\textwidth]{\includegraphics[width=0.325\textwidth]{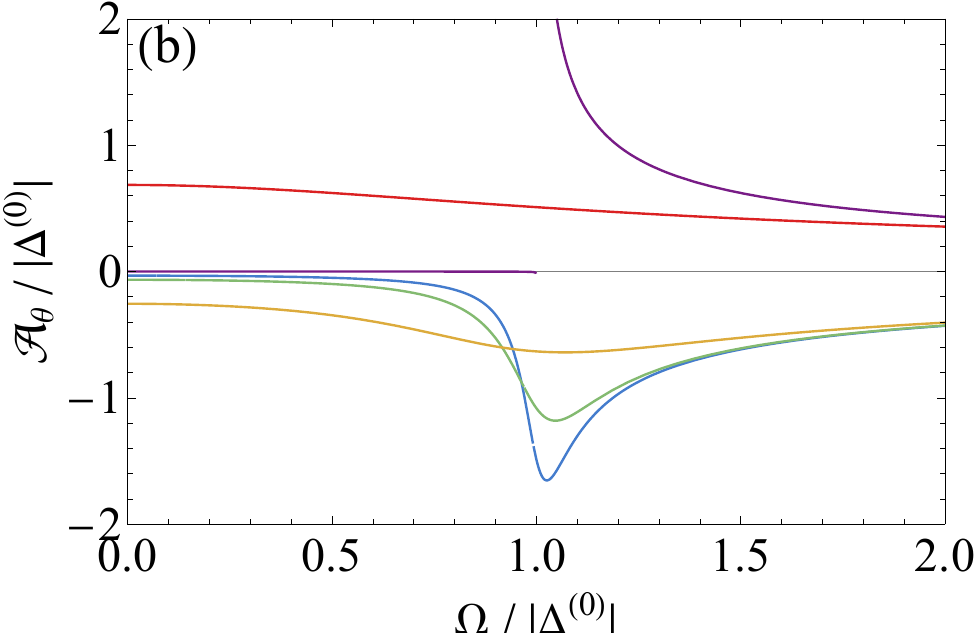}}\label{fig:TanPhiVaried}}
  \hfill
  {\makebox[0.325\textwidth]{\includegraphics[width=0.325\textwidth]{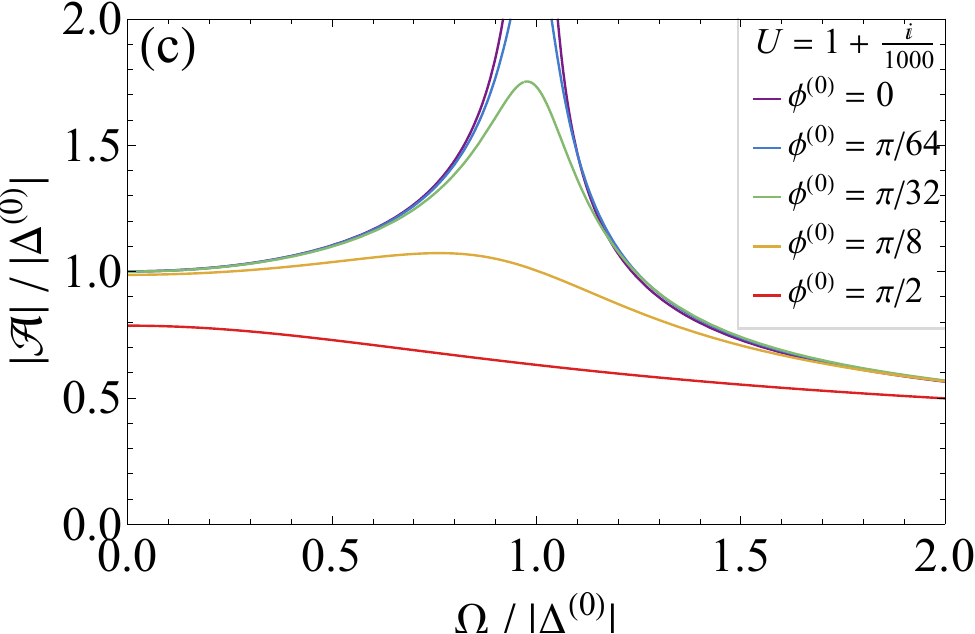}}\label{fig:NormPhiVaried}}
  \caption{(a) Radial component or Higgs mode, (b) tangential component or non-Hermitian phase mode, and (c) amplitude norm of the order parameter oscillation $\Delta^{(2)}(t)$ for different values of $\phi^{(0)}$ for complex pairing interaction $U = 1 + 10^{-3}\ i$.
  }\label{fig:PhiVaried}
\end{figure*}

Since $\mathbf{\sigma_k}^{(0)}$ is time independent,
the zeroth order equation, Eq. \eqref{eq:eom_zero}, reduces to $\mathbf{B_k}^{(0)} \times \boldsymbol{\sigma}_\mathbf{k}^{(0)} = 0$. This implies $\boldsymbol{\sigma}_\mathbf{k}^{(0)} =\lambda \mathbf{B_k}^{(0)}$ for $\lambda \in \mathbb{C}$. From the form of the Hamiltonian Eq. \eqref{eq:NH_bcs_pseudospin} we see that the real part of the energy is minimum when the pseudospin vectors $\boldsymbol{\sigma}_\mathbf{k}^{(0)}$ are anti-aligned with the complex pseudomagnetic field $\mathbf{B_k}^{(0)}$. This configuration corresponds to the effective ground state for a non-Hermitian Hailtonian \cite{Yamamoto2019}. Imposing the appropriate pseudospin normalization as done in \cite{Tsuji2015} gives the zero order pseudospin
\begin{equation}
    \label{eq:sigma0}
    \boldsymbol{\sigma}_\mathbf{k}^{(0)} = - \frac{\mathbf{B_k}^{(0)}}{2 |\mathbf{B_k}^{(0)}|} = - \frac{(\Delta^{(0)}, 0, \xi_\mathbf{k})}{2 \sqrt{|\Delta^{(0)}|^2 + |\xi_\mathbf{k}|^2}}
\end{equation}
subject to the self-consistency equation
\begin{equation}
    \label{eq:sce_zeroth}
    U \sum_\mathbf{k} \frac1{2 \sqrt{|\Delta^{(0)}|^2 + \xi_\mathbf{k}^2}} = 1.
\end{equation}

We solve the first order equation, Eq. \eqref{eq:eom_first}, using Fourier transform, which we detail in Appendix \ref{appendixsec:first_order}. Upon imposing the self-consistent equation, we find only trivial solutions
\begin{align}
    \label{eq:delta1}
    \Delta^{(1)}(t) = \bar{\Delta}^{(1)}(t) = 0.
\end{align}

The absence of first order corrections to the order parameters is expected since the leading contribution from external driving is second order as we have seen from Eq.~\eqref{eq:Bz_perturb}. 

Similarly, the second order equation, Eq. \eqref{eq:eom_second}, is solved using Fourier transform procedure (see Appendix \ref{appendixsec:second_order} for details). Imposing the self-consistent equations \eqref{eq:nscemetricized1a} and \eqref{eq:nscemetricized1b}, we find the second order corrections to be equal
\begin{equation}
    \label{eq:Fdelta2_condition}
    \Delta^{(2)}(t) = \bar{\Delta}^{(2)}(t).
\end{equation} 

In the Hermitian limit, this condition reduces to $\Delta^{(2)}(t)=\Delta^{(2)}(t)^*$, implying that $\Delta^{(2)}(t)$ is purely real and therefore corresponds to the Higgs mode in the gauge $\theta^{(0)}=0$. In the non-Hermitian case, however, $\bar{\Delta}\neq \Delta^*$ in general, so the second-order correction is generally complex. Furthermore, Eq.~\eqref{eq:Fdelta2_condition} shows that $\Delta^{(2)}$ and $\bar{\Delta}^{(2)}$ oscillate in phase, in contrast to the out-of-phase oscillations of $\Delta=\Delta^{(0)}e^{i\delta\theta}$ and $\Delta^\dagger=\Delta^{(0)}e^{-i\delta\theta}$ in the Hermitian case, which correspond to the Nambu-Goldstone mode. We therefore see that the oscillations in the phase $\phi$ are fundamentally different from the Nambu-Goldstone phase oscillations $\delta\theta$.

The explicit form of the second order correction is 
\begin{equation}
    \label{eq:delta2cos}
    \frac{\Delta^{(2)}(t)}{\frac{1}{2} |\mathbf{A}_0|^2 \alpha_1 /N(\varepsilon_F)} \sim \mathcal{A} \cos(2 \Omega t),
\end{equation}
where the complex amplitude $\mathcal{A}$ is given by
\begin{align}
    \label{eq:amplitude}
    \mathcal{A} =& \Delta^{(0)} \bigg[ U \sqrt{-1 + \frac{1}{1 + \frac{\Omega^2}{|\Delta^{(0)}|^2}- \exp(2i \phi^{(0)})}} \nonumber \\ 
    &\text{ArcCot}\! \left( \sqrt{-1 + \frac{1}{1 + \frac{\Omega^2}{|\Delta^{(0)}|^2}- \exp(2i \phi^{(0)})}}\right) \bigg]^{-1}.    
\end{align}

Physically, this corresponds to an oscillatory steady state of the order parameter. Notably, the order parameter oscillates at twice the driving frequency, $2\Omega$, consistent with the fact that this is a second order correction. We also note that the driving field in Eq.~\eqref{eq:driving} is assumed to be present for all times, from $t=-\infty$ to $t=+\infty$. For a quenched system, in which the external drive is switched on at (t=0), one may instead employ the Laplace transformation rather than the Fourier transformation. In that case, we expect behavior analogous to Eq.~\eqref{eq:delta2cos} in the long-time regime $t \gg 1/\Delta^{(0)}$, which is the time-scale of transient changes in the order parameter \cite{Yuzbashyan2006A, Yuzbashyan2006B}. However, this must also be reconciled with the assumption that the non-Hermitian BCS model that we consider is only applicable for time-scales much less than the inverse dissipation rate $1/(2\gamma)$. Thus, Eq. \eqref{eq:delta2cos} is applicable for time scales
\begin{equation}
    \label{eq:timescale}
    1/|\Delta^{(0)}| \ll t \ll 1/\gamma \;\; \;\text{or} \;\;\; |\Delta^{(0)}| \gg \gamma .    
\end{equation}
In terms of the characteristic cutoff frequency, which is assumed $\omega_C \gg |\Delta^{(0)}|$ in the BCS treatment, the condition is $\omega_C \gg \gamma$.

\begin{figure*}[!ht]
    \centering
    \includegraphics[width=\textwidth]{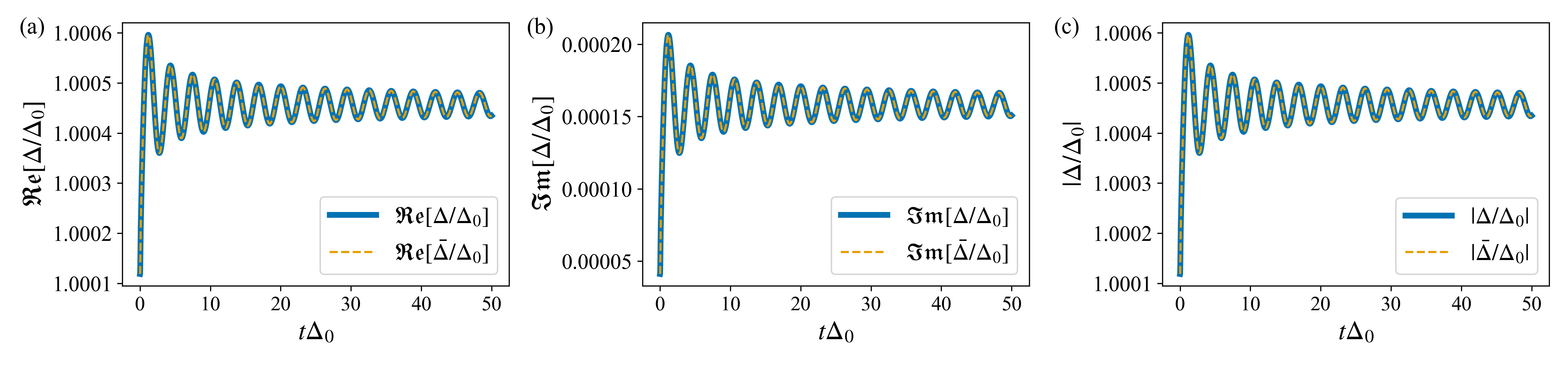}
    \caption{(a) Real part, (b) imaginary part, and (c) norm of the order parameters $\Delta(t)$ (solid blue line) and $\bar{\Delta}(t)$ (dashed orange line) following a quench at $t=0$ from a purely real interaction to a complex interaction with a finite imaginary component. The parameters are chosen as bandwidth $W/\Delta_0 = 46.8$ and $2\gamma/\Delta_0 = 10^{-3}$, consistent with Eq. \eqref{eq:timescale}. The two curves overlap throughout the time evolution, demonstrating that $\Delta(t)=\bar{\Delta}(t)$. The dynamics settles to an oscillation about a mean value given by Eq. \eqref{eq:complex_delta0}.
}
    \label{fig:quench}
\end{figure*}

The amplitude $\mathcal{A}$ is, in general, complex, since both the pairing interaction $U = U_0 + i\gamma$ and the equilibrium order parameter $\Delta^{(0)} = |\Delta^{(0)}| e^{i\phi^{(0)}}$ are complex. This is in contrast to the purely real amplitude oscillation in the Hermitian limit of $U \in \mathbb{R}$ and $\phi^{(0)} = 0$. Indeed, substituting these values to Eq. \eqref{eq:amplitude} yields a purely real amplitude for $\Omega < \Delta^{(0)}$, with a resonance at driving frequency $\Omega = \Delta^{(0)}$, consistent with previous studies \cite{Tsuji2015}. Physically, this corresponds to the Higgs mode, which is a radial oscillation in the complex order parameter space, and the absence of a Nambu-Goldstone phase oscillation, at least to second order, in the Hermitian case.

Meanwhile, in the non-Hermitian case, the complex amplitude acquires both radial and tangential components in the complex $\Delta^{(0)}$ plane. This leads to two possible interpretations. One is to regard the Higgs mode itself as becoming complex, with distinct real and imaginary parts. The other is to retain the definition of the Higgs mode as the radial oscillation in the complex order-parameter space. Here, we adopt the latter viewpoint, as it has the advantage of cleanly separating radial fluctuations—which reduce to the conventional Higgs mode in the Hermitian limit—from phase fluctuations, which are uniquely associated with the non-Hermiticity of the system. However, we emphasize that this phase oscillation is distinct from the conventional Nambu–Goldstone mode, which corresponds to oscillations of the gauge phase $\theta$. We therefore refer to this excitation as a non-Hermitian phase mode, to distinguish it from the conventional Goldstone mode.


\section{Resonance}
\label{sec:resonance}

Let us now look at the resonance of the Higgs and non-Hermitian phase modes in detail. We are interested in these collective modes as functions of the driving frequency $\Omega$. Figures \ref{fig:PhiVaried} (a), (b), and (c) show the Higgs mode, the non-Hermitian phase mode, and the norm of the scaled amplitude $\mathcal{A}/|\Delta^{(0)}|$, respectively, for various values of $\phi^{(0)}$. For these plots, we set the energy scale $|\Delta^{(0)}|=1$ and $U = 1 + 10^{-3}\ i$, which is consistent with the condition Eq. \eqref{eq:timescale}.

Figure~\ref{fig:PhiVaried}(a) shows that the radial amplitude (Higgs mode) in the complex order parameter space diverges at $\Omega = |\Delta^{(0)}|$, consistent with the behavior in the Hermitian limit reported in Ref.~\cite{Tsuji2015}. We emphasize that the system remains non-Hermitian even when $\phi^{(0)}=0$, since the interaction $U$ is complex; accordingly, the same divergence at $\Omega = |\Delta^{(0)}|$ persists in this case. As $\phi^{(0)}$ is increased to finite values, the divergence is regularized into finite peaks (blue and green curves), which are progressively suppressed and ultimately becomes flat at $\phi^{(0)}=\pi/2$. This indicates that the Higgs mode resonance is absent when the zeroth-order order parameter $\Delta^{(0)}$ is purely imaginary.

Figure~\ref{fig:PhiVaried}(b) displays the non-Hermitian phase oscillations for various values of $\phi^{(0)}$. For $\phi^{(0)}=0$, the response remains zero for $\Omega < |\Delta^{(0)}|$, in contrast to the Higgs oscillations shown in Fig.~\ref{fig:PhiVaried}(a). Otherwise, the behavior closely parallels that of the Higgs mode. In particular, there is a divergence at $\Omega = |\Delta^{(0)}|$. As $\phi^{(0)}$ increases to finite values, this resonance is also regularized into finite peaks (blue and green curves), which are subsequently suppressed and ultimately vanish at $\phi^{(0)}=\pi/2$.

Interestingly, even in the absence of the Anderson–Higgs mechanism, the non-Hermitian phase mode is gapped, as evidenced by the resonance peak at a nonzero driving frequency $\Omega=|\Delta_0|$ in Fig.~\ref{fig:PhiVaried}(b). This implies that the mode is gapped even in neutral fermionic superfluids, in sharp contrast to conventional BCS theory, where a gap in the phase mode emerges only for charged Cooper pairs through the Anderson-Higgs mechanism. Although we have referred to the radial fluctuation as the Higgs mode, this result suggests that the non-Hermitian phase fluctuation may also be interpreted as the phase component of a \textit{complex Higgs} mode in a non-Hermitian superfluid, which inherits its gapped character.

Finally, Fig.~\ref{fig:PhiVaried}(c) shows that, at fixed $U$, the resonance peak in the amplitude norm broadens and decreases with increasing $\phi^{(0)}$. The resonances associated with both the Higgs mode and the non-Hermitian phase mode disappear at $\phi^{(0)}=\pi/2$, corresponding to a purely imaginary $\Delta^{(0)}$, as illustrated by the red curves in the three panels of Fig.~\ref{fig:PhiVaried}. Interestingly, this suppression of the collective modes coincides with the emergence of exceptional points discussed in Sec.~\ref{subsec:exceptionalpoints}. It remains an open question whether such suppression is a generic feature of non-Hermitian superfluids hosting exceptional points near the fermi surface where the Cooper pairing mostly occur.

\section{Quench Dynamics}
\label{sec:quench}

To further illustrate the pseudospin dynamics of a non-Hermitian fermionic superfluid, we also consider the dynamics induced by an instantaneous interaction quench from a real interaction $U_0$ to a complex-valued interaction $U = U_0 + i \gamma$. Physically, this means instantaneously introducing a dissipation of $2 \gamma$ to the system. Consequently, the system becomes non-Hermitian, characterized by two order parameters, $\Delta$ and $\bar{\Delta}$, with dynamics given by equations of motion Eq. \eqref{eq:eom2} and self-consistent equations Eqs. \eqref{eq:scemetricized2a} and \eqref{eq:scemetricized2b}. We solve these numerically using 4th order Runge-Kutta for $N=10^4$ equally-spaced energy modes $\xi \in (-W/2, W/2)$ within a bandwidth $W/\Delta_0 = 46.8$ of the Fermi energy and with constant density of states, $\nu = 1/W$ \cite{Barankov2006, Yamamoto2021}.

The initial state is a Hermitian BCS ground state with interaction strength $U_0 \in \mathbb{R}$ and pseudospin configuration given by Eq. \eqref{eq:sigma0} but with $\Delta^{(0)} \in \mathbb{R}$. In the BCS approximation, the self-consistent equation relates $\Delta_0, W, \nu, \text{ and }U_0$ by $\Delta_0~=~W~e^{-1/\nu U_0}$.

Meanwhile, for $t>0$, the equilibrium order parameter $\Delta_0' \in \mathbb{C}$ for the complex interaction strength $U = U_0 + i \gamma$ is given by \cite{Yamamoto2019}
\begin{align}
    \label{eq:complex_delta0}
    \Delta_0' = \frac{W/2}{\sinh \left( 1/{\nu U}\right)}.
\end{align}
Upon the complex-interaction quench at $t=0$, both $\Delta(t)$ and $\bar{\Delta}(t)$ exhibit oscillatory dynamics shown in Fig.~\ref{fig:quench}. We find that the real and imaginary parts of $\Delta(t)$ and $\bar{\Delta}(t)$ oscillate in phase, implying that $\Delta(t)=\bar{\Delta}(t)$ throughout the evolution. This behavior is consistent with the analytical condition derived in Eq.~\eqref{eq:Fdelta2_condition}. We also observe that the dynamics decays to an oscillation about a numerical mean value approximately given by Eq. \eqref{eq:complex_delta0}.

Consequently, the phases of the two order parameters also oscillate in phase. This behavior is in sharp contrast to the conventional Nambu--Goldstone mode, characterized by $\Delta(t)=|\Delta_0|e^{i\theta(t)}$ and $\Delta^\dagger(t)=|\Delta_0|e^{-i\theta(t)}$, where the phases oscillate in antiphase. Our numerical results therefore provide further evidence for the emergence of a non-Hermitian phase mode in fermionic superfluids with complex interactions, which is fundamentally distinct from the conventional Nambu-Goldstone mode.

\section{Conclusions}
We investigated the collective excitations of fermionic superfluids described by a non-Hermitian BCS Hamiltonian with complex-valued interactions. Since the original biorthogonal formulation is not well suited for describing dynamics and time-dependent perturbations, we reformulated the theory in terms of pseudospins within the metricized framework. This approach provides a consistent extension of Heisenberg-picture dynamics to non-Hermitian systems, from which we derived the pseudospin equations of motion together with the self-consistency equations.

Using a perturbative analysis of the resulting semiclassical equations, we obtained analytical expressions for the collective modes in the presence of an external sinusoidal drive. In addition to the conventional Higgs mode, we identified a non-Hermitian phase mode that is fundamentally distinct from the Nambu-Goldstone mode.

These collective excitations exhibit several features absent in conventional Hermitian superfluids. In particular, the non-Hermitian phase mode remains gapped even in neutral superfluids, without requiring the Anderson--Higgs mechanism. Furthermore, the amplitudes of both the radial and phase oscillations depend explicitly on the equilibrium complex order parameter, $\Delta^{(0)}$, whose imaginary component originates from non-Hermiticity. Consequently, when $\mathrm{Im},\Delta^{(0)} \neq 0$, the resonant response remains finite, in contrast to the divergent behavior characteristic of the Hermitian limit.

Our results demonstrate that non-Hermiticity qualitatively modifies the collective dynamics of fermionic superfluids and gives rise to phase excitations beyond the conventional framework of equilibrium BCS theory.

\appendix
\section{First Order Solution}
\label{appendixsec:first_order}
The Fourier transform of the first order equation, Eq. \eqref{eq:eom_first}, of motion in index notation is
\begin{align}
    -i\omega \tilde{\sigma}^{\mu(1)}_\mathbf{k}(\omega)= 2\epsilon^{\mu \nu \rho} (B^{\nu (0)}_\mathbf{k} \tilde{\sigma}^{\rho (1)}_\mathbf{k}(\omega) + \tilde{B}^{\nu (1)}_\mathbf{k}(\omega)\sigma^{\rho (0)}_\mathbf{k}) ,
\end{align}
where $\epsilon^{\mu \nu \rho}$ is the Levi-Civita symbol. This can be rearranged and written in matrix form
\begin{align}
    \label{eq:Fsigma1}
    \tilde{\boldsymbol{\sigma}}^{(1)}_\mathbf{k}(\omega) = (-i\omega \mathbbm{1} - 2[\textbf{B}^{(0)}_\mathbf{k}]_\times)^{-1} 2\tilde{\textbf{B}}_\mathbf{k}^{(1)}(\omega) \times \boldsymbol{\sigma}^{(0)}_\mathbf{k} ,
\end{align}
where $[\textbf{B}^{(0)}_\mathbf{k}]_\times$ is the matrix representation of the cross product operator such that $([\textbf{B}^{(0)}_\mathbf{k}]_\times)^{\mu \rho}=\epsilon^{\mu \nu \rho} B^{\nu (0)}_\mathbf{k}$. This yields $\tilde{\boldsymbol{\sigma}}^{(1)}_\mathbf{k}$ in terms of the variables $\omega, \xi_\mathbf{k}, \Delta, \text{and } \bar{\Delta}$. Then, imposing the self-consistency equations yields the matrix equation
\begin{align}
    \label{eq:sce_first}
    \begin{pmatrix}
        A_{11}(\omega) & A_{12}(\omega)\\
        A_{21}(\omega) & A_{22}(\omega)
    \end{pmatrix}
    \begin{pmatrix}
        \tilde{\Delta}^{(1)}(\omega) \\
        \tilde{\bar{\Delta}}^{(1)}(\omega)
    \end{pmatrix}
    = \begin{pmatrix}
        0 \\
        0
    \end{pmatrix},
\end{align}
where $\tilde{\Delta}^{(1)}(\omega):= \mathcal{F}\{ \Delta^{(1)}(t) \}$, $\tilde{\bar{\Delta}}^{(1)}(\omega):= \mathcal{F}\{ \bar{\Delta}^{(1)}(t) \}$, and the coefficients are
\begin{align}
    A_{11}(\omega) &= \frac{1}{2} - U (\omega f_1(\omega) +2 f_2(\omega))\\
    A_{12}(\omega) &= \frac{1}{2} + U (\omega f_1(\omega) - 2 f_2(\omega))\\
    A_{21}(\omega) &= -\frac{1}{2} + U [2 (\Delta^{(0)})^2 f_0(\omega) + \omega f_1(\omega) + 2 f_2(\omega) ]\\ 
    A_{22}(\omega) &= \frac{1}{2} - U [2 (\Delta^{(0)})^2 f_0(\omega) - \omega f_1(\omega) + 2 f_2(\omega) ],
\end{align}
where
\begin{align}
    \label{eq:sumk_0}
    f_0(\omega) &= \sum_\mathbf{k} \frac{1}{2 \sqrt{|\Delta^{(0)}|^2+|\xi_\mathbf{k}|^2} \left(4 (\Delta^{(0)})^2 - \omega^2 + 4 \xi_\mathbf{k}^2 \right)}\\
    \label{eq:sumk_1}
    f_1(\omega) &= \sum_\mathbf{k} \frac{\xi_\mathbf{k}}{2 \sqrt{|\Delta^{(0)}|^2+|\xi_\mathbf{k}|^2} \left(4 (\Delta^{(0)})^2 - \omega^2 + 4 \xi_\mathbf{k}^2 \right)}\\
    \label{eq:sumk_2}
    f_2(\omega) &= \sum_\mathbf{k} \frac{\xi_\mathbf{k}^2}{2 \sqrt{|\Delta^{(0)}|^2+|\xi_\mathbf{k}|^2} \left(4 (\Delta^{(0)})^2 - \omega^2 + 4 \xi_\mathbf{k}^2 \right)} .
\end{align}

In the thermodynamic limit and for constant density of states, these sums can be expressed integrals that evaluate to
\begin{align}
    \label{eq:integrals}
    f_0(\omega) &= \frac{N(\varepsilon_F)}{4 |\Delta^{(0)}|^2} \frac1{1-a} \frac1{\sqrt{\frac{a}{1-a}}} \text{ArcTan}\!\left( \frac1{\sqrt{\frac{a}{1-a}}}\right) \\
    f_1(\omega) &= 0 \\
    f_2(\omega) &=\frac{1}{4U} - \frac{N(\varepsilon_F)}{4}\,\frac{a}{1-a} \frac{1}{\sqrt{\frac{a}{1-a}}} \text{ArcTan}\!\left( \sqrt{\frac{1-a}{a}} \right),
\end{align}
where
\begin{align}
    \label{eq:a_constant}
    a \equiv e^{i 2 \phi^{(0)}} - \frac{\omega^2}{4 |\Delta^{(0)}|^2} .    
\end{align}

We note that Eq. \eqref{eq:sce_first} only has non-trivial solutions for $\tilde{\Delta}^{(1)}(\omega), \tilde{\bar{\Delta}}^{(1)}(\omega)$ if the determinant of the left-hand side matrix is zero. No such non-trivial solution exists for arbitrary frequency $\omega$. Thus, we conclude that the only physical solutions are
\begin{align}
    \tilde{\Delta}^{(1)}(\omega) = \tilde{\bar{\Delta}}^{(1)}(\omega) = 0 ,
\end{align}
which implies the first order correction to the order parameters are also zero, and thus do not contribute to a collective oscillation, i.e.
\begin{align}
    \label{eq:delta1}
    \Delta^{(1)}(t) = \bar{\Delta}^{(1)}(t) = 0.
\end{align}
The absence of a first order correction to the order parameters is to be expected since from Eq. \eqref{eq:Bz_perturb}, the leading contribution from external driving is a second order correction.

\section{Second Order Solution}
\label{appendixsec:second_order}
The Fourier transform of the second order equation, Eq. \eqref{eq:eom_second}, yields the matrix equation
\begin{equation}
    \label{eq:Fsigma2}
    \tilde{\boldsymbol{\sigma}}_\mathbf{k}^{(2)}(\omega) = (-i \omega \mathbbm{1} - 2 [\textbf{B}^{(0)}_\mathbf{k}]_\times)^{-1}\ 2 \tilde{\textbf{B}}^{(2)}_\mathbf{k}(\omega) \times \ \boldsymbol{\sigma}^{(0)}_\mathbf{k} ,
\end{equation}
where we note that the $\mathbf{B_k}^{(1)}(t) \times \boldsymbol{\sigma}_\mathbf{k}^{(1)}(t)$ term vanished since substituting Eq. \eqref{eq:delta1} to Eqs. \eqref{eq:Bx_perturb},  \eqref{eq:By_perturb}, and \eqref{eq:Bz_perturb} yields $\mathbf{B_k}^{(1)}(t)=0$. Imposing the self-consistency equations Eqs. \eqref{eq:nscemetricized1a} and \eqref{eq:nscemetricized1b} to the second order pseudospin yields
\begin{align}
    \label{eq:sce_second}
    \begin{pmatrix}
        A_{11}(\omega) & A_{12}(\omega)\\
        A_{21}(\omega) & A_{22}(\omega)
    \end{pmatrix}
    \begin{pmatrix}
        \tilde{\Delta}^{(2)}(\omega) \\
        \tilde{\bar{\Delta}}^{(2)}(\omega)
    \end{pmatrix}
    = \begin{pmatrix}
        A_1(\omega) \\
        A_2(\omega)
    \end{pmatrix} ,
\end{align}
where
\begin{align}
    A_{11}(\omega) =& \frac{1}{2} - U (\omega f_1(\omega) + 2 f_2(\omega))\\
    A_{12}(\omega) =& \frac{1}{2} + U (\omega f_1(\omega) - 2 f_2(\omega))\\
    A_{1}(\omega) =& \sqrt{2 \pi} |\mathbf{A}_0|^2 U \alpha_1 \Delta^{(0)} f_2(\omega) [\delta(\omega + 2 \Omega)\nonumber\\
    &- 2 \delta(\omega) + \delta(\omega - 2 \Omega)]\\
    A_{21}(\omega) =& -\frac{1}{2} + U [2 (\Delta^{(0)})^2 f_0(\omega) + \omega f_1(\omega) + 2 f_2(\omega) ]\\ 
    A_{22}(\omega) =& \frac{1}{2} - U [2 (\Delta^{(0)})^2 f_0(\omega) - \omega f_1(\omega) + 2 f_2(\omega) ] \\
    A_{2}(\omega) =& \sqrt{2 \pi} |\mathbf{A}_0|^2 U \alpha_1 \Delta^{(0)} f_1(\omega) \left[\frac{1}{2}\delta(\omega + 2 \Omega) -  \delta(\omega)\right.\nonumber\\
    &\left.+ \frac{1}{2} \delta(\omega - 2 \Omega) \right].
\end{align}
where $f_0(\omega), f_1(\omega), \text{ and } f_2(\omega)$ are as defined in Eqs. \eqref{eq:sumk_0}, \eqref{eq:sumk_1}, and \eqref{eq:sumk_2}. We note that the RHS of Eq. \eqref{eq:sce_second} is non-zero, in contrast to that for first-order Eq. \eqref{eq:sce_first}, hence we expect a non-zero second order correction for the order parameters. 

From Eq. \eqref{eq:integrals}, $f_1(\omega) = 0$ and the second row equality in Eq. \eqref{eq:sce_second} reduces to:
\begin{equation}
    \label{eq:Fdelta2_condition_appendix}
    \tilde{\Delta}^{(2)}(\omega) = \tilde{\bar{\Delta}}^{(2)}(\omega) ,
\end{equation} 
Substituting Eq. \eqref{eq:Fdelta2_condition_appendix} to Eq. \eqref{eq:sce_second} yields an expression for $\tilde{\Delta}^{(2)}(\omega)$, which upon performing inverse Fourier transform gives the second-order correction to the order parameters 
\begin{align}
    \label{eq:delta2}
    \frac{\Delta^{(2)}(t)}{\frac1{2}|\mathbf{A}_0|^2 \alpha_1} = \Delta^{(0)} \bigg(\frac{4U\ f_2(0)}{-1 + 4U\,f_2(0)}\ 
    + \frac{4U\ f_2(2\Omega) \cos(2 \Omega t)}{1 - 4U\,f_2(2\Omega)}\bigg) .
\end{align}
For characterizing the resonance of the system in response to external driving, the relevant term is the second term inside the parentheses in Eq. \eqref{eq:delta2}. Further simplification using Eq. \eqref{eq:integrals} yields Eq. \eqref{eq:delta2cos}.


\bibliography{apssamp}

@PREAMBLE{
 "\providecommand{\noopsort}[1]{}" 
 # "\providecommand{\singleletter}[1]{#1}%" 
}

@article{Yamamoto2019,
  title = {Theory of Non-Hermitian Fermionic Superfluidity with a Complex-Valued Interaction},
  author = {Yamamoto, Kazuki and Nakagawa, Masaya and Adachi, Kyosuke and Takasan, Kazuaki and Ueda, Masahito and Kawakami, Norio},
  journal = {Phys. Rev. Lett.},
  volume = {123},
  issue = {12},
  pages = {123601},
  numpages = {7},
  year = {2019},
  month = {Sep},
  publisher = {American Physical Society},
  doi = {10.1103/PhysRevLett.123.123601},
  url = {https://link.aps.org/doi/10.1103/PhysRevLett.123.123601}
}

@article{Ghatak2018,
  title = {Theory of superconductivity with non-Hermitian and parity-time reversal symmetric Cooper pairing symmetry},
  author = {Ghatak, Ananya and Das, Tanmoy},
  journal = {Phys. Rev. B},
  volume = {97},
  issue = {1},
  pages = {014512},
  numpages = {14},
  year = {2018},
  month = {Jan},
  publisher = {American Physical Society},
  doi = {10.1103/PhysRevB.97.014512},
  url = {https://link.aps.org/doi/10.1103/PhysRevB.97.014512}
}

@article{Tsuji2015,
  title = {Theory of Anderson pseudospin resonance with Higgs mode in superconductors},
  author = {Tsuji, Naoto and Aoki, Hideo},
  journal = {Phys. Rev. B},
  volume = {92},
  issue = {6},
  pages = {064508},
  numpages = {11},
  year = {2015},
  month = {Aug},
  publisher = {American Physical Society},
  doi = {10.1103/PhysRevB.92.064508},
  url = {https://link.aps.org/doi/10.1103/PhysRevB.92.064508}
}

@article{Ju2025,
  title = {Heisenberg and Heisenberg-like representations via Hilbert-space-bundle geometry in the non-Hermitian regime},
  author = {Ju, Chia-Yi and Miranowicz, Adam and Barnett, Jacob and Chen, Guang-Yin and Nori, Franco},
  journal = {Phys. Rev. A},
  volume = {111},
  issue = {5},
  pages = {052213},
  numpages = {7},
  year = {2025},
  month = {May},
  publisher = {American Physical Society},
  doi = {10.1103/PhysRevA.111.052213},
  url = {https://link.aps.org/doi/10.1103/PhysRevA.111.052213}
}

@article{Ju2019,
  title = {Non-Hermitian Hamiltonians and no-go theorems in quantum information},
  author = {Ju, Chia-Yi and Miranowicz, Adam and Chen, Guang-Yin and Nori, Franco},
  journal = {Phys. Rev. A},
  volume = {100},
  issue = {6},
  pages = {062118},
  numpages = {16},
  year = {2019},
  month = {Dec},
  publisher = {American Physical Society},
  doi = {10.1103/PhysRevA.100.062118},
  url = {https://link.aps.org/doi/10.1103/PhysRevA.100.062118}
}

@article{Sim2025,
  title = {Observables in non-Hermitian systems: A methodological comparison},
  author = {Sim, Karin and Defenu, Nicol\`o and Molignini, Paolo and Chitra, R.},
  journal = {Phys. Rev. Res.},
  volume = {7},
  issue = {1},
  pages = {013325},
  numpages = {11},
  year = {2025},
  month = {Mar},
  publisher = {American Physical Society},
  doi = {10.1103/PhysRevResearch.7.013325},
  url = {https://link.aps.org/doi/10.1103/PhysRevResearch.7.013325}
}

@article{HeissSteeb1991,
  author  = {Heiss, W. D. and Steeb, W.-H.},
  title   = {Analytical properties of {H}amiltonian eigenvalues and exceptional points},
  journal = {Journal of Mathematical Physics},
  volume  = {32},
  number  = {11},
  pages   = {3003--3007},
  year    = {1991},
  doi     = {10.1063/1.529699}
}

@article{Heiss2012,
  author  = {Heiss, W. D.},
  title   = {Exceptional points of non-{H}ermitian operators},
  journal = {Journal of Physics A: Mathematical and Theoretical},
  volume  = {45},
  number  = {44},
  pages   = {444016},
  year    = {2012},
  doi     = {10.1088/1751-8113/45/44/444016}
}

@article{Dembowski2001,
  author  = {Dembowski, C. and Gr{\"a}f, H.-D. and Harney, H. L. and Heine, A. and Heiss, W. D. and Rehfeld, H. and Richter, A.},
  title   = {Experimental observation of the topological structure of exceptional points},
  journal = {Physical Review Letters},
  volume  = {86},
  number  = {5},
  pages   = {787--790},
  year    = {2001},
  doi     = {10.1103/PhysRevLett.86.787}
}

@article{Gong2018,
  author  = {Gong, Zongping and Ashida, Yuto and Kawabata, Kohei and Takasan, Kazuaki and Higashikawa, Seiji and Ueda, Masahito},
  title   = {Topological phases of non-{H}ermitian systems},
  journal = {Physical Review X},
  volume  = {8},
  number  = {3},
  pages   = {031079},
  year    = {2018},
  doi     = {10.1103/PhysRevX.8.031079}
}

@article{Lin2020,
  author  = {Lin, Rongrong and Xiao, Li and Deng, Tie and Wang, Le and Yi, Wei and Xue, Ping},
  title   = {Topological non-{H}ermitian skin effect},
  journal = {Physical Review Letters},
  volume  = {125},
  number  = {14},
  pages   = {146802},
  year    = {2020},
  doi     = {10.1103/PhysRevLett.125.146802}
}

@article{Takasu2020,
  author  = {Takasu, Yuta and Ono, Akira and Hirano, Tetsuya and Takahashi, Yoshiro},
  title   = {{$\mathcal{PT}$}-symmetric non-{H}ermitian quantum many-body system using ultracold atoms},
  journal = {Physical Review Letters},
  volume  = {124},
  number  = {13},
  pages   = {130405},
  year    = {2020},
  doi     = {10.1103/PhysRevLett.124.130405}
}

@article{Chen2022,
  author  = {Chen, Tianyu and Li, Yao and Ren, Hongchao and Zhang, Zhenhua and Chen, Wenjing and Yan, Zhiyuan and Xue, Ping},
  title   = {Quantum {Z}eno effects across a parity-time symmetry breaking transition in atomic momentum space},
  journal = {Physical Review Letters},
  volume  = {128},
  number  = {18},
  pages   = {180403},
  year    = {2022},
  doi     = {10.1103/PhysRevLett.128.180403}
}

@article{Schindler2011,
  author  = {Schindler, Joseph and Li, Ang and Zheng, Mei C. and Ellis, F. M. and Kottos, Tsampikos},
  title   = {Experimental study of active {LRC} circuits with {$\mathcal{PT}$} symmetries},
  journal = {Physical Review A},
  volume  = {84},
  number  = {4},
  pages   = {040101(R)},
  year    = {2011},
  doi     = {10.1103/PhysRevA.84.040101}
}

@article{Helbig2020,
  author  = {Helbig, Tobias and Hofmann, Tobias and Imhof, Steffen and Abdelghany, Marwa and Kiessling, Tobias and Molenkamp, Laurens W. and Lee, Ching Hua and Szameit, Alexander and Greiter, Martin and Thomale, Ronny},
  title   = {Generalized bulk--boundary correspondence in non-{H}ermitian topolectrical circuits},
  journal = {Nature Physics},
  volume  = {16},
  number  = {7},
  pages   = {747--750},
  year    = {2020},
  doi     = {10.1038/s41567-020-0922-9}
}

@article{Naghiloo2018,
  author  = {Naghiloo, M. and Jordan, A. N. and Murch, K. W.},
  title   = {Information gain and loss in a continuously monitored qubit},
  journal = {Physical Review Letters},
  volume  = {121},
  number  = {3},
  pages   = {030405},
  year    = {2018},
  doi     = {10.1103/PhysRevLett.121.030405}
}

@article{Ashida2020,
  author  = {Ashida, Yuto and Gong, Zongping and Ueda, Masahito},
  title   = {Non-{H}ermitian physics},
  journal = {Advances in Physics},
  volume  = {69},
  number  = {3},
  pages   = {249--435},
  year    = {2020},
  doi     = {10.1080/00018732.2020.1876991}
}

@article{Deng2010,
  author  = {Deng, Hui and Haug, Hartmut and Yamamoto, Yoshihisa},
  title   = {Exciton-polariton {B}ose--{E}instein condensation},
  journal = {Reviews of Modern Physics},
  volume  = {82},
  number  = {2},
  pages   = {1489--1537},
  year    = {2010},
  doi     = {10.1103/RevModPhys.82.1489}
}

@article{Gao2015,
  author  = {Gao, Tian and Estrecho, E. and Bliokh, K. Y. and Liew, T. C. H. and Fraser, M. D. and Brodbeck, S. and Kamp, M. and Schneider, C. and Yamamoto, Y. and Nori, F. and Ostrovskaya, E. A. and Kavokin, A. V.},
  title   = {Observation of non-{H}ermitian degeneracies in a chaotic exciton-polariton billiard},
  journal = {Nature},
  volume  = {526},
  number  = {7574},
  pages   = {554--558},
  year    = {2015},
  doi     = {10.1038/nature15522}
}

@article{Tamura2025,
  author  = {Tamura, Shun and M{\"u}ller, Helene and Aliani, Linus and Kornich, Viktoriia},
  title   = {Meissner effect in non-{H}ermitian superconductors},
  journal = {Physical Review B},
  volume  = {111},
  number  = {18},
  pages   = {L180503},
  year    = {2025},
  doi     = {10.1103/PhysRevB.111.L180503}
}

@article{Hoinka2017Goldstone,
  author  = {Hoinka, Sascha and Dyke, Paul and Lingham, Marcus G. and Kinnunen, Jami J. and Bruun, Georg M. and Vale, Chris J.},
  title   = {Goldstone mode and pair-breaking excitations in atomic Fermi superfluids},
  journal = {Nature Physics},
  volume  = {13},
  number  = {10},
  pages   = {943--946},
  year    = {2017},
  doi     = {10.1038/nphys4187}
}

@article{Uchino2020NGPointContact,
  author  = {Uchino, Shun},
  title   = {Role of Nambu-Goldstone modes in the fermionic-superfluid point contact},
  journal = {Physical Review Research},
  volume  = {2},
  number  = {2},
  pages   = {023340},
  year    = {2020},
  doi     = {10.1103/PhysRevResearch.2.023340}
}

@article{Measson2014,
  title = {Amplitude Higgs mode in the $2H\ensuremath{-}{\text{NbSe}}_{2}$ superconductor},
  author = {M\'easson, M.-A. and Gallais, Y. and Cazayous, M. and Clair, B. and Rodi\`ere, P. and Cario, L. and Sacuto, A.},
  journal = {Phys. Rev. B},
  volume = {89},
  issue = {6},
  pages = {060503},
  numpages = {5},
  year = {2014},
  month = {Feb},
  publisher = {American Physical Society},
  doi = {10.1103/PhysRevB.89.060503},
  url = {https://link.aps.org/doi/10.1103/PhysRevB.89.060503}
}

@article{Matsunaga2013,
  title = {Higgs Amplitude Mode in the BCS Superconductors ${\mathrm{Nb}}_{1\mathrm{\text{\ensuremath{-}}}x}{\mathrm{Ti}}_{x}\mathbf{N}$ Induced by Terahertz Pulse Excitation},
  author = {Matsunaga, Ryusuke and Hamada, Yuki I. and Makise, Kazumasa and Uzawa, Yoshinori and Terai, Hirotaka and Wang, Zhen and Shimano, Ryo},
  journal = {Phys. Rev. Lett.},
  volume = {111},
  issue = {5},
  pages = {057002},
  numpages = {5},
  year = {2013},
  month = {Jul},
  publisher = {American Physical Society},
  doi = {10.1103/PhysRevLett.111.057002},
  url = {https://link.aps.org/doi/10.1103/PhysRevLett.111.057002}
}

@article{Behrle2018,
  title = {Higgs mode in a strongly interacting fermionic superfluid},
  author = {Behrle, A. and Harrison, T. and Kombe, J. and Gao, K. and Link, M. and Bernier, J.-S. and Kollath, C. and Köhl, M.},
  journal = {Nature Physics},
  volume = {14},
  issue = {},
  pages = {781-785},
  numpages = {5},
  year = {2018},
  month = {},
  publisher = {Nature},
  doi = {https://doi.org/10.1038/s41567-018-0128-6},
  url = {https://www.nature.com/articles/s41567-018-0128-6}
}

@article{Vidanovic2014,
  title = {Dissipation through localized loss in bosonic systems with long-range interactions},
  author = {Vidanovi\ifmmode \acute{c}\else \'{c}\fi{}, Ivana and Cocks, Daniel and Hofstetter, Walter},
  journal = {Phys. Rev. A},
  volume = {89},
  issue = {5},
  pages = {053614},
  numpages = {9},
  year = {2014},
  month = {May},
  publisher = {American Physical Society},
  doi = {10.1103/PhysRevA.89.053614},
  url = {https://link.aps.org/doi/10.1103/PhysRevA.89.053614}
}

@article{Barontini2013,
  title = {Controlling the Dynamics of an Open Many-Body Quantum System with Localized Dissipation},
  author = {Barontini, G. and Labouvie, R. and Stubenrauch, F. and Vogler, A. and Guarrera, V. and Ott, H.},
  journal = {Phys. Rev. Lett.},
  volume = {110},
  issue = {3},
  pages = {035302},
  numpages = {5},
  year = {2013},
  month = {Jan},
  publisher = {American Physical Society},
  doi = {10.1103/PhysRevLett.110.035302},
  url = {https://link.aps.org/doi/10.1103/PhysRevLett.110.035302}
}

@article{
    Tomita2017,
    author = {Takafumi Tomita  and Shuta Nakajima  and Ippei Danshita  and Yosuke Takasu  and Yoshiro Takahashi },
    title = {Observation of the Mott insulator to superfluid crossover of a driven-dissipative Bose-Hubbard system},
    journal = {Science Advances},
    volume = {3},
    number = {12},
    pages = {e1701513},
    year = {2017},
    doi = {10.1126/sciadv.1701513},
    URL = {https://www.science.org/doi/abs/10.1126/sciadv.1701513},
    eprint = {https://www.science.org/doi/pdf/10.1126/sciadv.1701513}
}

@article{Yamamoto2021,
  title = {Collective Excitations and Nonequilibrium Phase Transition in Dissipative Fermionic Superfluids},
  author = {Yamamoto, Kazuki and Nakagawa, Masaya and Tsuji, Naoto and Ueda, Masahito and Kawakami, Norio},
  journal = {Phys. Rev. Lett.},
  volume = {127},
  issue = {5},
  pages = {055301},
  numpages = {8},
  year = {2021},
  month = {Jul},
  publisher = {American Physical Society},
  doi = {10.1103/PhysRevLett.127.055301},
  url = {https://link.aps.org/doi/10.1103/PhysRevLett.127.055301}
}

@article{Yuzbashyan2006A,
  title = {Relaxation and Persistent Oscillations of the Order Parameter in Fermionic Condensates},
  author = {Yuzbashyan, Emil A. and Tsyplyatyev, Oleksandr and Altshuler, Boris L.},
  journal = {Phys. Rev. Lett.},
  volume = {96},
  issue = {9},
  pages = {097005},
  numpages = {4},
  year = {2006},
  month = {Mar},
  publisher = {American Physical Society},
  doi = {10.1103/PhysRevLett.96.097005},
  url = {https://link.aps.org/doi/10.1103/PhysRevLett.96.097005}
}

@article{Yuzbashyan2006B,
  title = {Dynamical Vanishing of the Order Parameter in a Fermionic Condensate},
  author = {Yuzbashyan, Emil A. and Dzero, Maxim},
  journal = {Phys. Rev. Lett.},
  volume = {96},
  issue = {23},
  pages = {230404},
  numpages = {4},
  year = {2006},
  month = {Jun},
  publisher = {American Physical Society},
  doi = {10.1103/PhysRevLett.96.230404},
  url = {https://link.aps.org/doi/10.1103/PhysRevLett.96.230404}
}

@article{Brody2014,
  title={Biorthogonal quantum mechanics},
  author={Brody, Dorje C},
  journal={Journal of Physics A: Mathematical and Theoretical},
  volume={47},
  number={3},
  pages={035305},
  year={2014},
  publisher={IOP Publishing}
}

@article{Mostafazadeh2004,
  title={Time-dependent Hilbert spaces, geometric phases, and general covariance in quantum mechanics},
  author={Mostafazadeh, Ali},
  journal={Physics Letters A},
  volume={320},
  number={5-6},
  pages={375--382},
  year={2004},
  publisher={Elsevier}
}

@article{Mostafazadeh2018,
  title = {Energy observable for a quantum system with a dynamical Hilbert space and a global geometric extension of quantum theory},
  author = {Mostafazadeh, Ali},
  journal = {Phys. Rev. D},
  volume = {98},
  issue = {4},
  pages = {046022},
  numpages = {18},
  year = {2018},
  month = {Aug},
  publisher = {American Physical Society},
  doi = {10.1103/PhysRevD.98.046022},
  url = {https://link.aps.org/doi/10.1103/PhysRevD.98.046022}
}

@article{Ghosh2022,
  title = {Non-Hermitian higher-order topological superconductors in two dimensions: Statics and dynamics},
  author = {Ghosh, Arnob Kumar and Nag, Tanay},
  journal = {Phys. Rev. B},
  volume = {106},
  issue = {14},
  pages = {L140303},
  numpages = {9},
  year = {2022},
  month = {Oct},
  publisher = {American Physical Society},
  doi = {10.1103/PhysRevB.106.L140303},
  url = {https://link.aps.org/doi/10.1103/PhysRevB.106.L140303}
}

@article{Ji2025,
  title = {Non-Hermitian second-order topological superconductors},
  author = {Ji, Xiang and Ding, Wenchen and Chen, Yuanping and Yang, Xiaosen},
  journal = {Phys. Rev. B},
  volume = {109},
  issue = {12},
  pages = {125420},
  numpages = {8},
  year = {2024},
  month = {Mar},
  publisher = {American Physical Society},
  doi = {10.1103/PhysRevB.109.125420},
  url = {https://link.aps.org/doi/10.1103/PhysRevB.109.125420}
}

@article{Barankov2006,
  title = {Synchronization in the BCS Pairing Dynamics as a Critical Phenomenon},
  author = {Barankov, R. A. and Levitov, L. S.},
  journal = {Phys. Rev. Lett.},
  volume = {96},
  issue = {23},
  pages = {230403},
  numpages = {4},
  year = {2006},
  month = {Jun},
  publisher = {American Physical Society},
  doi = {10.1103/PhysRevLett.96.230403},
  url = {https://link.aps.org/doi/10.1103/PhysRevLett.96.230403}
}


\end{document}